\documentclass[10pt,conference]{IEEEtran}
\usepackage{amsmath, amssymb, bm, cite, epsfig, psfrag}
\usepackage{graphicx}
\usepackage[top    = 0.75in,
bottom = 1in,
left   = 0.625in,
right  = 0.625in]{geometry}
\usepackage{dblfloatfix}
\usepackage{array}
\usepackage{textcomp}
\usepackage{float}
\usepackage{longtable}
\usepackage{supertabular,booktabs}
\usepackage{multirow}
\usepackage[usenames,dvipsnames]{xcolor}

\usepackage{etoolbox}
\usepackage{pbox}
\usepackage{fixltx2e}
\usepackage{romannum}
\usepackage{filecontents}
\usepackage{enumerate}
\graphicspath{ {figures/} }
\usepackage{subfig}
\usepackage{colortbl}
\usepackage{fancyhdr}

\usepackage{bm}
\newtoggle{conference}
\togglefalse{conference} %
\graphicspath{{figures/}}

\fancypagestyle{firststyle}{
	\fancyhf{}
	\fancyhead[L]{O. Kanhere and T. S. Rappaport, ``Calibration of NYURay, a 3D mmWave and sub-THz Ray Tracer using Indoor, Outdoor, and Factory Channel Measurements,'' \textit{in ICC 2023 - 2023 IEEE International Conference on Communications,} Rome, Italy, May 2023, pp. 1--6.}     %

}
\usepackage{etoolbox}
\makeatletter
\patchcmd{\@makecaption}
{\scshape}
{}
{}
{}
\makeatletter
\patchcmd{\@makecaption}
{\\}
{.\ }
{}
{}
\makeatother
\IEEEoverridecommandlockouts
\begin{document}
\title{Calibration of NYURay, a 3D mmWave and sub-THz Ray Tracer using Indoor, Outdoor, and Factory Channel Measurements}
\author{\IEEEauthorblockN{Ojas Kanhere, Theodore S. Rappaport \thanks{This work is supported by the NYU WIRELESS Industrial Affiliates Program and National Science Foundation (NSF) Grants: 1909206 and 2037845.}}
	
\IEEEauthorblockA{	\small NYU WIRELESS\\
					 Tandon School of Engineering, New York University\\
					Brooklyn, NY 11201\\
					\{ojask, tsr\}@nyu.edu}}

\maketitle
\thispagestyle{firststyle}

\begin{abstract}
Ray tracing is a powerful tool that can be used to predict wireless channel characteristics, reducing the need for extensive channel measurements for channel characterization, evaluation of performance of sensing applications such as position location, and wireless network deployment. In this work, NYURay, a 3D mmWave and sub-THz ray tracer, is introduced, which is calibrated to wireless channel propagation measurements conducted at 28, 73, and 140 GHz, in indoor office, outdoor, and factory environments. We present an accurate yet low-complexity calibration procedure to obtain electrical properties of materials in any environment by modeling the reflection coefficient of building materials to be independent of the angle of incidence, a simplification shown to be quite effective in \cite{Schaubach92a} over 30 years ago. We show that after calibration, NYURay can accurately predict individual directional multipath signal power. The standard deviation in the error of the directional multipath power predicted by the ray tracer compared to the directional measured power was less than 3 dB in indoor office environments and less than 2 dB in outdoor and factory environments.
\end{abstract}
    

\section{Introduction}\label{Introduction}

The vast spectrum at mmWave and sub-THz frequencies will enable a wide variety of new applications such as centimeter-level position location, high-resolution virtual/augmented reality, factory automation, environment sensing, and imaging \cite{Rappaport19a}. To determine the feasibility of such applications, tremendous efforts have been made to characterize the wireless channel. Numerous channel measurements have been conducted at mmWave and sub-THz frequencies to develop accurate channel models \cite{Ju20,Mac15b}. Real-world channel measurements however are time consuming, with typical measurement campaigns collecting at most tens of locations per environment. Additionally, conducting wireless channel measurements is often cost-prohibitive due to the complex hardware required. Site-specific simulation-based prediction of wireless channels can supplement actual measurements, if there is an assurance that signal characteristics of predicted wireless channels, such as the powers, angles of arrival and departure (AoA/AoD), and time-of-flight (ToF) of signals arriving at wireless receivers at a wide range of locations closely matches the measured signal characteristics at said locations. 

Ray tracing is a powerful tool that can be used to predict the signal propagation characteristics of wireless signals based on the location of the wireless transmitter (TX), receiver (RX), and obstructions in the environment. Obstructions are objects in the environment that may give rise to reflections, transmissions or scattering when impinged by a propagating signal. The power, AoA, AoD, and ToF of the wireless signals are predicted, which can then be used to reproduce the spatial and temporal characteristics of the wireless system for simulation-based analysis. 

The evaluation of the performance of sensing applications such as position location and seeing through walls can be supplemented using ray tracing. Synthetic data can be generated to evaluate the performance of the sensing algorithms at locations where the channel measurements were not conducted \cite{Kanhere19a}.


Utilizing a ray tracer for wireless simulations ``out of the box" without calibration from field measurements leads to incorrect predictions of the wireless channel. For example, a ray tracer that predicts greater spatial diversity in the wireless channel could lead to an overestimation of the channel capacity. Inaccurate power predicted by the ray tracer, if used for network planning, could lead to sub-optimal network base station placement resulting in regions of signal outage.

This paper presents the theoretical propagation models used by NYURay, a 3-D mmWave and sub-THz ray tracer developed at NYU WIRELESS,  to simulate reflection, penetration and scattering of wireless signals for site-specific wireless channel prediction. A description of the procedure that can be used to quickly calibrate NYURay to direction channel measurements conducted in a variety of indoor and outdoor environments at 28, 73, and 142 GHz is provided (over a total of nine different measurement campaigns from 2012 to 2022 \cite{Rappaport19a, Mac15b, Rap13a, Xing_2021c, Ju_2023,Shakya_2023}), along with performance results of utilizing the calibration procedure for prediction of signal propagation. The calibration procedure results in a standard deviation of error in predicted received power in a directional setup of less than 3 dB even when only a rudimentary environmental map is available. Additionally, the electrical properties of building materials, as obtained from ray tracer calibration, are provided.

\section{Prior mmWave and sub-THz Ray Tracers and Metrics for Calibration}
A variety of research groups and commercial entities have developed mmWave and sub-Thz ray tracers in indoor and outdoor environments. To evaluate the performance of ray tracers, channel statistics, such as path loss, angular spread, and delay spread are often used for comparison. In \cite{Lee_2018}, authors compared the measured and predicted omnidirectional path loss in an Urban Microcell (UMi) environment at 28 GHz. Researchers fine-tuned a single value of relative permittivity of materials in the environment to minimize the error in path loss prediction in an indoor office environment \cite{Liu_2017} at 28 GHz. The power-angular-delay profile of the measured and simulated environment was compared for a large empty room and a small office at 29 GHz (for a single TX-RX location pair in both environments) in \cite{Karstensen_2016}. The dominant specular paths were shown to be predicted by the ray tracer, however a quantitative study of the difference in the measured channel and simulated channel was not done. In \cite{Zhou_2017_RT}, the path gains of individual multipath components (MPCs) were compared at 60 GHz in an indoor office room. However, a direct comparison of simulated directional power to the measured power is rarely done. The material parameters up to 100 GHz were extrapolated to sub-THz frequencies in \cite{Gougeon_2019}. In \cite{Han_2015}, the authors measured the refractive index of material such as wallpaper and plaster via THz time-domain spectroscopy and calculated the reflection coefficients via Fresnel's equations, observing good agreement. However, the reflection and scattering parameters were directly used for ray tracer, without validation in more complex environments. Kurner et al. \cite{Kurner_2012} calibrated an indoor sub-THz ray tracer at 300 GHz to directional measurements conducted at TX-RX locations by minimizing the difference in power of the measured and simulated single-bounce rays. In \cite{Guan_2019b}, the authors characterized the wireless channel inside a train wagon at 60 and 300 GHz, and calibrated an in-house ray tracer to the path gain, delay, and AoA/AoD of significant measurement rays. Ray tracing simulations in a corridor were validated for LOS locations in \cite{Sheikh_2022} at 90, 95, and 100 GHz by comparing the measured and simulated RX power. 

\section{Ray Propagation Mechanisms}

Signals propagating through a wireless media interact with objects in the environment, resulting in changes in the direction of propagation and the power carried by the signal. At mmWave and sub-THz frequencies, reflection and transmission are the dominant propagation mechanisms observed in field measurements, with diffraction being negligible at these high frequencies \cite{Kurner_2012}. In regions where specular reflections from the TX are blocked, diffused scattering is also a significant propagation mechanism. NYURay supports specular reflections, penetration, and diffuse scattering while ignoring diffraction.

In the literature, reflection loss is often modeled using Fresnel's equations \cite{Rap02a}. However, due to the highly non-linear nature of Fresnel's equations, a closed form solution to minimizing the difference in the measured MPC powers to the simulated ray tracer output to field measurements is not attainable. In the absence of detailed knowledge of the complex environment where measurements were conducted, the measured reflection loss can not be accurately modeled via Fresnel's equations. As seen in \cite{Choi_2015}, no clear dependence between angle of incidence and reflected power was observed, since insufficient information of the material being tested (such as fine-grain details of the material under test) was available. We propose the use of a simplified and more analytically tractable reflection loss model, wherein the reflection loss of a ray impinging an obstruction is assumed to be constant, independent of incident angle. As shall be explained in Section \ref{sec:calibration}, such simplification allows for a closed-form solution for minimizing the difference in the measured and simulated path gains, allowing quick and accurate calibration of the electrical properties of the materials in the environment. The wall attenuation factor (WAF) model \cite{Durgin_1998} was used to model the signal partition loss through materials. To model the power carried by the scattering rays, the directive scattering model was employed, which was shown to match scattering patterns observed at mmWave and sub-THz frequencies in \cite{Xing_2021d,Ju19a}. 

 NYURay uses a hybrid ray tracing algorithm \cite{Tan96a}, which combines the shooting-bouncing rays (SBR) ray tracing technique \cite{Schaubach92a, Durgin97a, Seidel94a} with geometry-based ray tracing \cite{McKown91a,Ho94a}. Reflections up to fifth bounce reflections along a propagation path were considered for ray tracing simulations, since higher order reflections were not observed in the field measurements at 28, 73, and 142 GHz \cite{Rap13a,Rappaport19a}.

\section{MmWave and sub-THz Channel Measurements}\label{sec:measurements}

MmWave and sub-THz channel measurements were conducted at 28, 73, and 140 GHz, in and around the Brooklyn and Manhattan campuses of NYU \cite{Rap13a,MacCartney_2019,Xing_2021c, Xing_2021d}, in the indoor office environment of the 9$ ^{th} $ floor of 2 Metrotech, Brooklyn  \cite{Mac15b,Xing_2021d,Xing19a}, and in factory environments \cite{Ju_2022b,Ju_2023}. The measurements conducted by graduate students at NYU WIRELESS, over a decade from 2011 to 2022, have led to the development of channel models for mmWave and sub-THz communications, proving the viability of mmWave and sub-THz for future communications \cite{Rap13a,Rappaport19a}. In this section, the hardware used to conduct the measurements is described. Additionally, the environments where the mmWave and sub-THz measurements were conducted are described. 

\subsection{Measurement equipment}\label{sec:equipment}
Three generations of channel sounders were developed at NYU WIRELESS, starting with the first sliding-correlation based channel sounder operating at 28 GHz, progressing to the mmWave channel sounder at 73 GHz, and the latest sub-THz channel sounder at 140 GHz, operated by many generations of graduate students from 2011 through 2022. All three channel sounders had similar super-heterodyne RF architectures \cite{Xing_2021d}. The RF signal was transmitted and received out of a high-gain directional horn antenna mounted on a rotatable gimbal.

To capture all arriving multipath propagation in the environment, the TX and RX gimbals were rotated in the azimuth and elevation. A channel power-delay-profile (PDP) was captured at different TX-RX antenna pointing combinations. The channel sounder used for propagation measurements at 28 GHz had an RF bandwidth of 800 MHz, while the channel sounders at 73 and 142 GHz utilized an RF bandwidth of 1 GHz.

\subsection{Measurement Campaigns and Environments}

Indoor office measurements were conducted at 28 at 142 GHz in an indoor open-office on the 9th floor of 2 Metrotech Center in 2014 and 2019, respectively \cite{Mac15b,Rap13a,Ju20}. The environment consisted of cubicles, private offices, classrooms, glass doors, elevators, and drywall walls. A map of the TX and RX measurement locations is available in \cite{Ju20}.

Outdoor measurements were conducted at 28, 73, and 142 GHz \cite{Rap13a,MacCartney_2019,Xing_2021c, Xing_2021d}. The 28 GHz measurements were conducted at the NYU campus in downtown Manhattan, which was an urban environment surrounded by general university areas and high rise buildings \cite{Rap13a}. The 73 and 142 GHz outdoor measurements were conducted in an open square environment on the NYU Tandon campus in downtown Brooklyn \cite{MacCartney_2019,Xing_2021d, Xing_2021c}. The TX-RX locations used in the 28, 73, and 140 GHz measurement campaign can be found in \cite{Rap13a}, \cite{MacCartney_2019}, and \cite{Xing_2021d}, respectively.  Cherry trees were in the center of the environment, surrounded by benches, lampposts, and buildings.

Four factory measurement campaigns were conducted at 140 GHz \cite{Ju_2022b,Ju_2023}. The first factory was two-storied, large factory building where multiple companies shared office space. The factory had open-seating areas, dedicated meeting rooms with glass panes and metallic frames, lockers, and metal elevators. The second factory was a medium-sized electronics manufacturing facility. The third factory was a medium sized warehouse facility. The warehouse had several metallic storage aisles stocked with cardboard boxes, and a package inspection room consisting of multiple desks for package inspection. The fourth factory was a single room manufacturing space. Prototyping machines were present in the environment. A map of the measurement locations is available in \cite{Ju_2022b,Ju_2023}

\section{Calibrating NYURay to Real-World Measurements}\label{sec:calibration}

The wide range of materials present in the diverse measurement environments, such as drywall, glass, granite, wooden cupboards, foliage, and metallic factory equipment, had differing electrical properties. To generate high-fidelity prediction of the measured wireless channel, it is critical for a ray tracer to accurately model the varying electrical properties of the materials in the environment. The directional channel propagation measurements conducted in the indoor, outdoor and factory environments at mmWave and sub-THz frequencies may be used to calibrate the ray tracer, to best fit the measured channel. 

To calibrate NYURay, we shall now introduce an approach that calibrates power of individual directional MPCs.

Consider a single MPC $ j $ arriving at the RX after traversing a distance of $ d $ m. NYURay models the power carried by the MPC ($ P_{j,R} $) as follows:
\begin{align}
	&P_{j,R} [dBm] = P_{j,TX} [dBm] +G_{j,T} [dBi] +G_{j,R} [dBi] \nonumber\\
	&-FSPL(d_j,f) [dB]-\sum_{i=1}^{N}\left(w^{pen}_{i,j}L^{pen}_{i,j}+ w^{ref}_{i,j}L^{ref}_{i,j}\right) [dB], \label{eq:pl}
\end{align}
where $ P_{j,TX} $ is the RF power at the transmitter (TX) in dBm, $  G_{j,T} $ and $ G_{j,R} $ are the antenna gains (in dBi) in the direction of departure and arrival of the signal at the TX and RX respectively, $ d_j $ is the distance traveled by the signal the TX to the RX in meters, $ FSPL(d_j,f) $ is the free space path loss (in dB) at a distance $ d_j $ and a frequency $ f $ (GHz). $ N $ is the number of material types in the environment. $w^{pen}_{i,j} L^{pen}_{i} $ is the total power lost (in dB) due to penetration loss of MPC $ j $ due to obstructions of material type $ i $. The weights $ w^{pen}_{i,j} $ and $  w^{ref}_{i,j}  $are set to be equal the number of times the multipath penetrates and reflects through obstacles with material type $ i $, respectively. $ L^{pen}_{i} $ and $L^{ref}_{i} $ are the power lost (in dB) at each such penetration, and reflection, respectively. 

To calibrate the ray tracer, the power measured in the strongest MPC of a directional measurement, where directional antennas were used at TX and RX as described above, was compared to the power predicted via \eqref{eq:pl}. If the power contained in the strongest MPC is given by $ P_{j,meas} $, the following sum of squares loss function was use to minimize the error between the measured MPC power and the model \eqref{eq:pl}:
\begin{align}
	L=	&\sum_{j=1}^{M} \left( P_{j,meas}-P_{j,R}\right)^2 \label{eq:LS},
\end{align}
where all quantities are in log-scale (e.g. dB or dBm). $ L $ is the squared loss function that is to be minimized and $ M $ is the number of MPCs over which calibration is done. The MPCs measured were obtained from multiple directional measurements at each RX location. 

Assuming the propagation path of the MPC is predicted by the ray tracer, \eqref{eq:LS} may be written as:
\begin{align}
	L=  \sum_{j=1}^{M}\left(A_j-\mathbf{w_j}\mathbf{L}\right)^2  \label{eq:LS2}
\end{align}
where $ \mathbf{w_j} = [w^{pen}_{1,j} w^{pen}_{2,j} \dotsc w^{pen}_{N,j} w^{ref}_{1,j} w^{ref}_{2,j} \dotsc w^{ref}_{N,j} ]$ and $ \mathbf{L} = [L^{pen}_{1} L^{pen}_{2} \dotsc L^{pen}_{N} L^{ref}_{1} L^{ref}_{2} \dotsc L^{ref}_{N}]^T  $, and $ A_j = P_{j,TX} +G_{j,T}  +G_{j,R} 
-FSPL(d_j,f) - P_{j,meas} $.

The optimal values of $ L^{pen}_{i} $and $ L^{ref}_{i} $ ($\hat{L}$) to minimize \eqref{eq:LS} are given by the least squares solution to the following set of linear equations:
\begin{align}
	\mathbf{W} \mathbf{L} = \mathbf{A},
\end{align}
where $ W = [	\mathbf{w_1}		\mathbf{	 w_2}  \dotsc		\mathbf{  w_N}]^T $.

Thus, $ \hat{L} = (W^TW)^{-1} W^T A$.


\section{ Performance of NYURay for Predicting mmWave and sub-THz Measurements }

\begin{figure}[]
	\centering
	\includegraphics[width=0.45\textwidth]{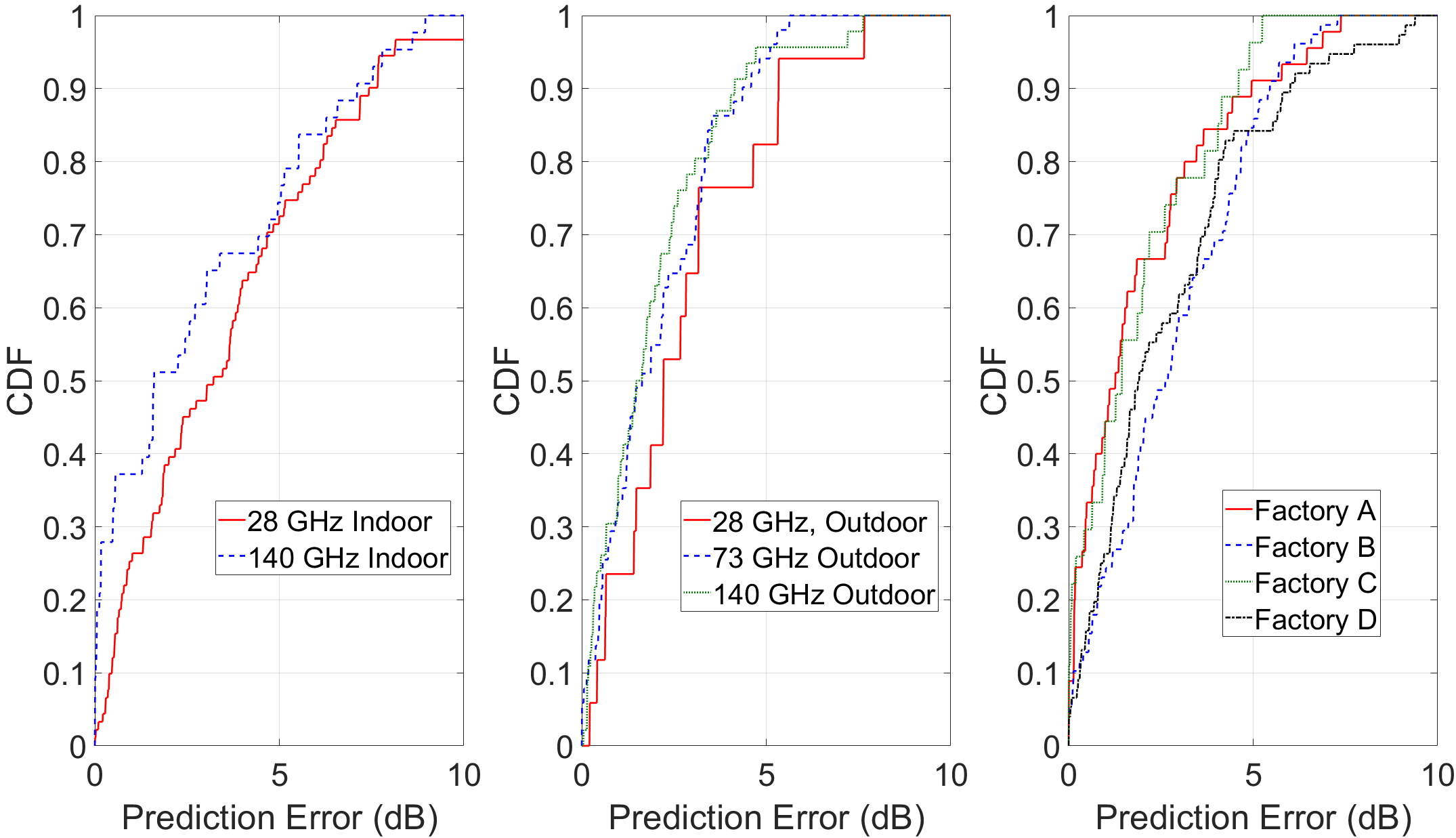}
	\caption{The absolute error in the multipath power predicted by NYURay ( $ |L| $ ) compared to the measured multipath power, over the nine measurement campaigns using (1)-(4).}
	\label{fig:power_error}
\end{figure}

Good agreement between the measured and simulated multipath powers was observed for the nine measurement campaigns, as seen in Fig. \ref{fig:power_error}. The difference in measured and predicted individual multipath power was minimized via least squares. For the indoor office measurement campaigns at 28 and 140 GHz, a mean error of 0.11 and 0.73 dB and a standard deviation of 2.7 dB and 2.8 dB, respectively, was obtained. NYURay was able to better predict the directional multipath power in less cluttered outdoor environments, with a mean power error in a directional MPCs being of 1.18 dB, 0.01 dB, and -0.01 dB and standard deviations of error equal to 2.0 dB, 1.6 dB and 1.7 dB at 28, 73, and 140 GHz, respectively. For factory environment measurements conducted at 140 GHz \cite{Ju_2023}, the detailed 360-degree videos of the environment enabled calibration to greater accuracy than the indoor office measurement campaigns, resulting in a mean error of individual directional multipath powers of 0.54 dB, 0.04 dB, 0.02 dB, and 0.2 dB, with a standard deviations of 2.0 dB, 1.9 dB, 1.7 dB, and 2.3 dB respectively at Factory A, B, C, and D.

\begin{figure}[]
	\centering
	\includegraphics[width=0.45\textwidth]{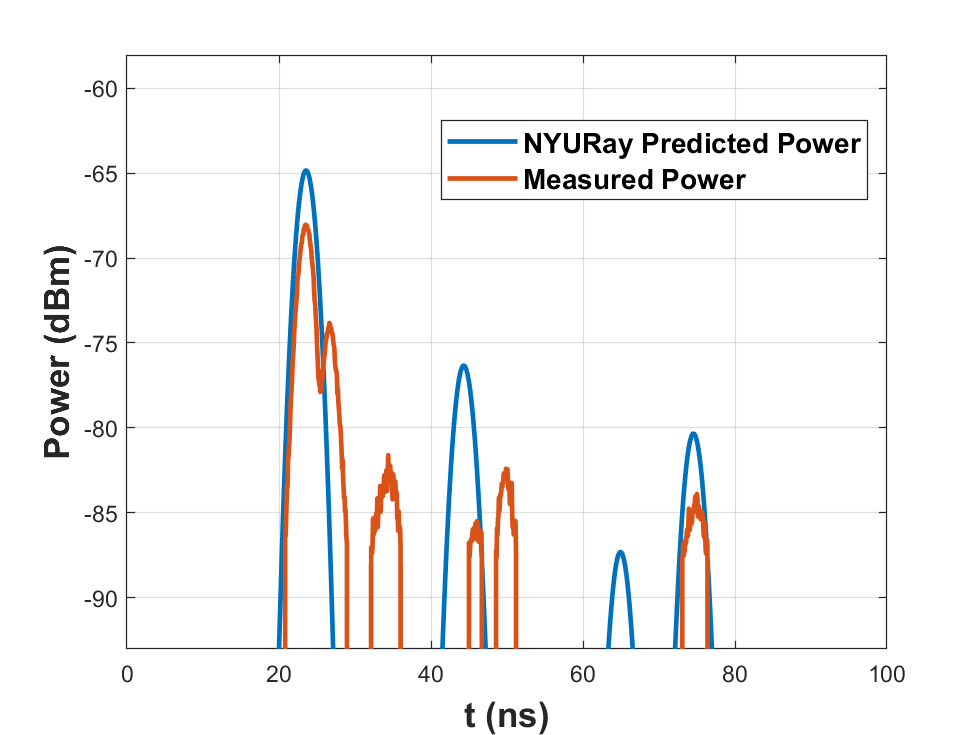}
	\caption{Comparison between the measured and simulated PDP obtained at an indoor location at 28 GHz after calibration using (1)-(4).}
	\label{fig:directional_pdp}
\end{figure}

%
A sample comparison of the PDP generated by NYURay and the PDP measured in the 28 GHz indoor office measurement campaign after calibration (e.g. use of (1)-(4)) is shown in Fig. \ref{fig:directional_pdp}. Although some comparatively weaker multipath measured at a delay of 35 ns are not predicted by NYURay, good agreement in the powers of other MPCs is observed. In particular, the measured power of the strongest MPC is within 2 dB of the power predicted by NYURay.

In addition to comparing the directional multipath power (which was directly calibrated via least squares estimation), the performance of NYURay was evaluated by comparing measured and simulated secondary channel statistics, namely the RMS delay spread and the angular spread. 

\begin{figure}[]
	\centering
	\includegraphics[width=0.45\textwidth]{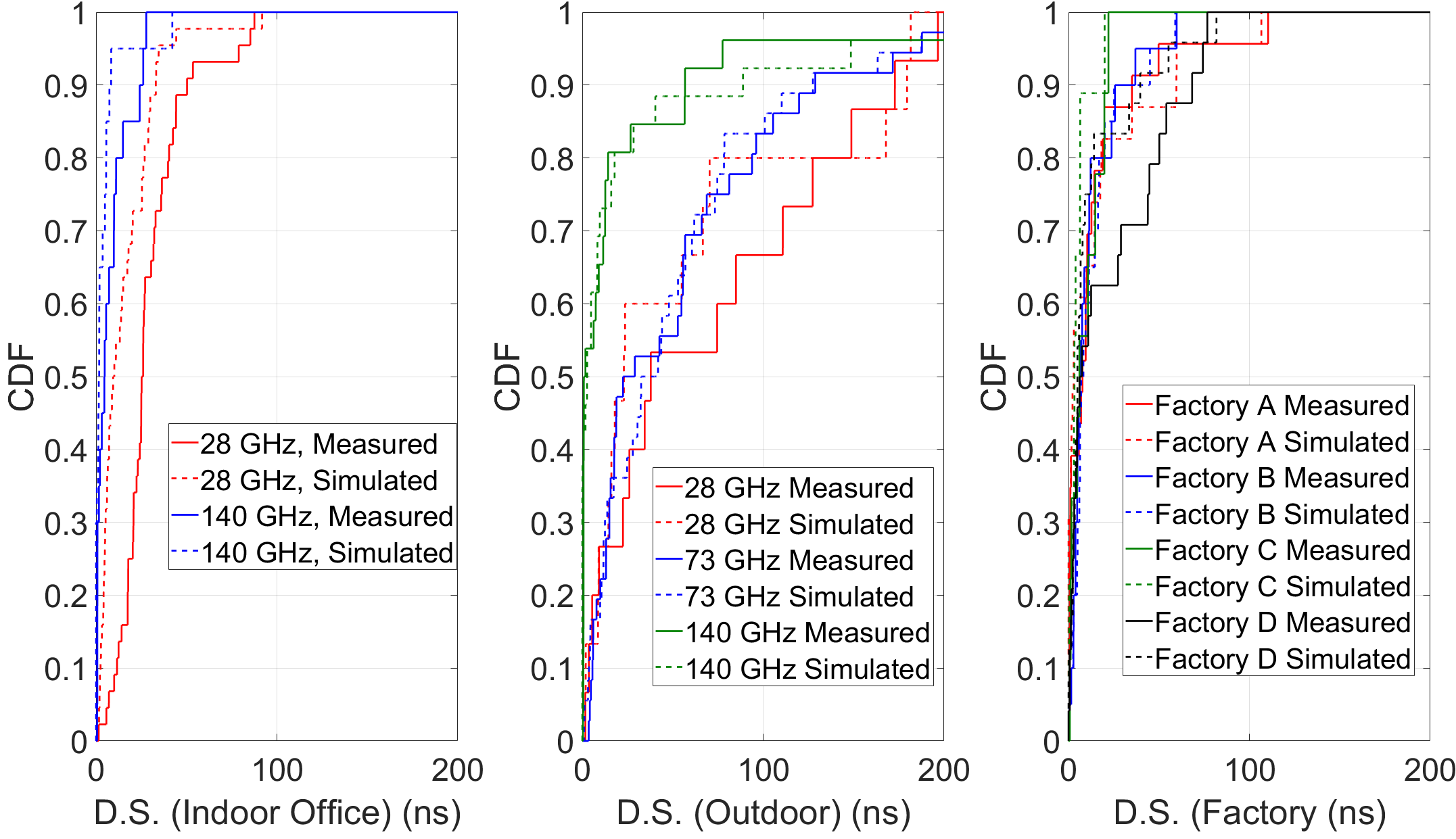}
	\caption{Comparison between the measured and simulated delay spread for the nine measurement campaigns.}
	\label{fig:delay_spread}
\end{figure}

\begin{figure}[]
	\centering
	\includegraphics[width=0.45\textwidth]{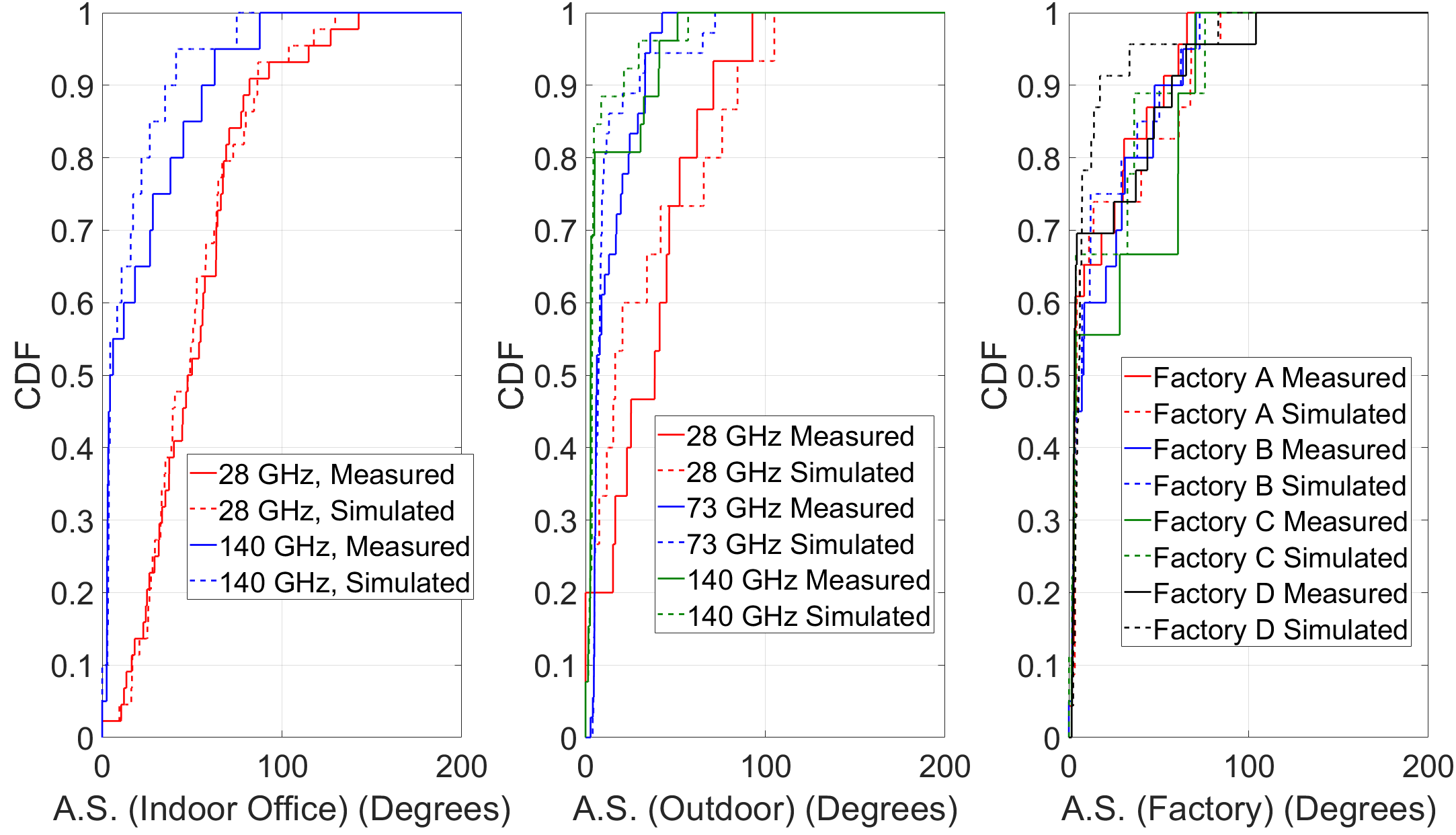}
	\caption{Comparison between the measured and simulated angular spread for the nine measurement campaigns.}
	\label{fig:angular_spread}
\end{figure}

As seen in Figs. \ref{fig:delay_spread} and \ref{fig:angular_spread}, good agreement was observed between measured and predicted angular and delay spreads after calibration. A lower delay spread was observed at higher frequencies due to greater penetration loss.

\section{Material Properties}

MmWave and sub-THz signals may be reflected, scattered, or may penetrate through materials in the environment. By characterizing reflection, transmission and scattering interactions of a variety of common building materials, signal propagation may be predicted to greater accuracy \cite{Ju19a, Seidel94a}. 

A wide range of materials were present in the diverse measurement environments, such as drywall, glass, granite, wooden cupboards, foliage, and metallic factory equipment. The penetration loss of objects was found to increase with increase in frequency. The reflection and penetration loss of drywall and glass obtained from ray tracer calibrations at 28 and 140 GHz agrees well with the measurements conducted in \cite{Olsson_2021}. Lower foliage loss was observed at 140 GHz compared to 73 GHz, possibly because the 140 GHz measurements were conducted in fall, when the trees had fewer leaves, while the 73 GHz measurements were conducted in summer. Materials were found to be most reflective at lower frequencies.

\begin{table}
	\begin{tabular}{|l|l|l|l|l|}
	\hline
	Material                    &\begin{tabular}[c]{@{}l@{}}Frequency\\ (GHz)\end{tabular} & Environment & \begin{tabular}[c]{@{}l@{}}Reflection \\ Loss (dB)\end{tabular} & \begin{tabular}[c]{@{}l@{}}Penetration \\ Loss (dB)\end{tabular} \\ \hline

	\multirow{6}{*}{Drywall}      &    28 & Indoor Office       & 6.1                                                       &4.0   \\ \cline{2-5}
	&    140 & Indoor Office       & 9.9                                                       &9.2   \\ \cline{2-5}
	& 140    &    Factory  A     & 8.7                                                        &13.1    \\ \cline{2-5}
	&140     &    Factory  B     & 12.8                                                        &12.7    \\\cline{2-5}
	&140    &    Factory  C     & 7.9                                                        &-    \\ \cline{2-5}
	&140    &    Factory  D     & 10.1	                                                        &8.0   \\ \hline
	
	\multirow{7}{*}{Glass}      &    28 & Indoor Office       & 3.5                                                       & 3.2   \\ \cline{2-5}
	&73&Outdoor&5.9 &5.0 \\ \cline{2-5}
	&    140 & Indoor Office       & 24.5                                                       & 7.2   \\ \cline{2-5}
	&140&Outdoor&7.4 &3.9 \\ \cline{2-5}
	& 140    &    Factory  A     & 9.9                                                        &10.4    \\ \cline{2-5}						
	&140    &    Factory  D     & 6.9                                                        &-    \\ \hline
	\pagebreak
	Thick Glass	& 140    &    Factory  A     & 8.4                                                       &23.0   \\ \hline						
	\multirow{2}{*}{	\begin{tabular}[c]{@{}l@{}}Cubicles\\(Fabric)\end{tabular} }      &    28 & Indoor Office       & 3.3                                                       &- \\ \cline{2-5}
	&    140 & Indoor Office       & 8.0                                                    &7.8   \\ \hline
	
	\multirow{2}{*}{	\begin{tabular}[c]{@{}l@{}}Wooden\\Cupboard\end{tabular} }      &    28 & Indoor Office       & 3.5                                                       &2.4   \\ \cline{2-5}
	&    140 & Indoor Office       & 0.5                                                       &6.1   \\ \hline				
	\multirow{2}{*}{Display Board}      &    28 & Indoor Office       & 1.1                                                       &11   \\ \cline{2-5}
	&    140 & Indoor Office       & 8.9                                                       &19.1   \\ \hline						
	\multirow{2}{*}{Whiteboard}      &    28 & Indoor Office       & 8.3                                                       &11   \\ \cline{2-5}
	&140    &    Factory  D     & -       &8.5    \\ \hline	
	Cardboard Box   &    140 & Factory C       & 4.1                                                       &1.7   \\ \hline
	Cork Board      
	&    140 & Factory C       &15.3 &-  \\ \hline   
	Wood & 140 &Factory D & 4.8 & -\\ \hline
	Cement Wall & 28 &Outdoor & 11.6&-\\ \hline
	Granite & 28 &Outdoor & 6.9&- \\ \cline{2-5}
	& 73 &Outdoor & 5.6&- \\ \cline{2-5}
	& 140 &Outdoor &13.1&- \\ \hline
	\pagebreak
	Concrete Pillar & 73 &Outdoor & 12.7&- \\ \cline{2-5}
	& 140 &Outdoor & 10.3&- \\ \hline
	\multirow{2}{*}{Brick Wall}   & 73&Outdoor&12.8&-\\ \cline{2-5}
	& 140&Outdoor&18.9&-\\ \hline
	\multirow{2}{*}{Foliage} &73&Outdoor&-&6.1\\ \cline{2-5}
	&140&Outdoor&-&4.6\\ \hline
	\end{tabular}
	\caption{Reflection and penetration loss of common building materials obtained from ray tracer calibration. The reflection and penetration loss increases with increase in frequency.}\label{tab:My_values}
\end{table}
\section{Conclusion}

This paper presented NYURay, a 3-D mmWave and sub-THz ray tracer calibrated to real-world channel measurements at 28, 73, and 140 GHz in indoor office, outdoor UMi, and factory environments. A calibration procedure was described, which assumed angle-independent reflection coefficients of building materials and minimized the loss of individual directional MPCs over many locations and environments. The calibration procedure was used to quickly calibrate a ray tracer to directional measurements, resulting in a standard deviation of error in the predicted multipath power of significant MPCs equal to 2.7 and 2.8 dB in the indoor office environment at 28 and 140 GHz, respectively, 2.0 dB, 1.6 dB and 1.7 dB in the outdoor environments at 28, 73, and 140 GHz. In the factory environments, the standard deviation in the error of the predicted multipath power was low (2.0 dB, 1.9 dB, 1.7 dB, and 2.3 dB in Factory A, B, C, and D). The angular spread and rms delay spread of omnidirectional measurements were under-predicted by about 20 \% in the indoor office and outdoor environments. This paper provided electrical properties of some common building materials, as determined from the ray tracer calibration procedure across 9 measurement campaigns from 2012 through 2022. The reflection and penetration loss increased with increase in frequency as shown in \cite{Xing19a, Rappaport19a}.

Ray tracing accuracy is improved with more accurate environmental maps, however we have shown that with a rudimentary environmental map containing major reflectors, quick and accurate ray tracer calibration is possible.

\end{document}